\documentclass[twocolumn,aps,prl]{revtex4}
\usepackage{amsmath} 
\usepackage{amssymb} 
\usepackage{graphicx} 
\usepackage{epsfig}
\usepackage{color}
\usepackage{epstopdf}
 \usepackage{rotating}
 
\begin{document}
 \title{Generating Weyl semimetals from alkali metals}
 \author{Joseba Goikoetxea, Jorge Bravo-Abad, and Jaime Merino}
\affiliation{Departamento de F\'isica Te\'orica de la Materia Condensada, Condensed Matter Physics Center (IFIMAC) and
Instituto Nicol\'as Cabrera, Universidad Aut\'onoma de Madrid, Madrid 28049, Spain}
\begin{abstract}
We report the discovery of a time-reversal symmetric Weyl semimetal obtained by modifying a model Hamiltonian describing the 
electronic properties of conventional alkali metals. The artificially generated Weyl semimetal features four isolated Weyl nodes in its bulk band structure 
and displays characteristic surface Fermi arcs arising from topologically protected surface states. The Weyl semimetal occurs
as an intermediate state between a conventional band insulator and a three-dimensional topological insulator.  
The generation of topological Weyl semimetals from conventional metals opens a new route towards the deterministic design of simple materials 
hosting Weyl fermions.
\end{abstract}
 \date{\today}
 \maketitle
 
{\em Introduction.}
Weyl fermions predicted around 90 years ago by H. Weyl \cite{weyl1929} have not yet been observed in high energy 
physics experiments. However, condensed matter systems have provided an alternative and versatile platform for the realization of Weyl fermions \cite{armitage2018}. In Weyl semimetals, 
despite the non-relativistic velocities of electrons, the crystal lattice potential leads to linear dispersions at the so-called Weyl points. This unique behavior is accompanied by the emergence 
of characteristic \emph{Fermi arcs}, connecting projected Weyl points at their surfaces and reflecting the topological origin of the corresponding surface bands. Remarkably, the non-trivial 
topological properties of Weyl semimetals have also enabled a variety of unusual physical phenomena such as the Adler-Bell-Jackiw chiral anomaly\cite{adler1969,bell1969} on a lattice\cite{spivak2013,dassarma2015}
and the anomalous quantum Hall effect\cite{luttinger1954,luttinger1958,fang2003}.

These interesting properties have stimulated an intense activity in searching for new materials hosting Weyl fermions. Fermi arcs have been observed in the family of TaAs
\cite{hasan2015,huang2015,weng2015,hasan2017}  materials which are Weyl semimetals with broken inversion symmetry. Weyl semimetals with broken time-reversal symmetry 
have been proposed in magnetically ordered pyrochlore iridates \cite{savrasov2011,balents2011,nakayama2016} and in
two-band insulators with an intrinsic chirality irradiated with linearly polarized light \cite{nagaosa2016}.  Despite the importance of these breakthroughs, the intrinsic accidental nature of band degeneracies leading to Weyl nodes\cite{herring1937}, makes the general endeavour of finding new Weyl semimetals an unsystematic and challenging task. One possibility to overcome this drawback is to generalize to simple electronic materials 
the band engineering techniques developed in the context of highly controllable bosonic structures, such as optical lattices\cite{chen2016,buljan2015,delplace2012,jiang2012} and photonic crystals\cite{soljacic2015,bravoabad2015}

In  this Letter we show how a Weyl semimetal can be generated from a conventional
alkali metal. By artificially modifying the hopping pattern entering the original tight-binding model, a Weyl 
semimetal which preserves time reversal symmetry but breaks spatial inversion 
symmetry is generated. This Weyl semimetal contains the minimum four band touching Weyl points 
with opposite chiralities in the Brillouin zone imposed by time reversal symmetry. Fermi arcs connecting the 
projection of the Weyl points arising from topological edge states are found on different crystal faces. 
The Weyl semimetal occurs as an intermediate state between a band insulator and
a three-dimensional topological insulator. Our proposal provides a route for generating Weyl nodes in conventional BCC 
metals as an alternative route to the tedious search of complex materials having accidental band touching points 
in the presence of broken inversion and/or time reversal symmetry.

\begin{figure}[t]
   \centering
    \includegraphics[width=8.5cm]{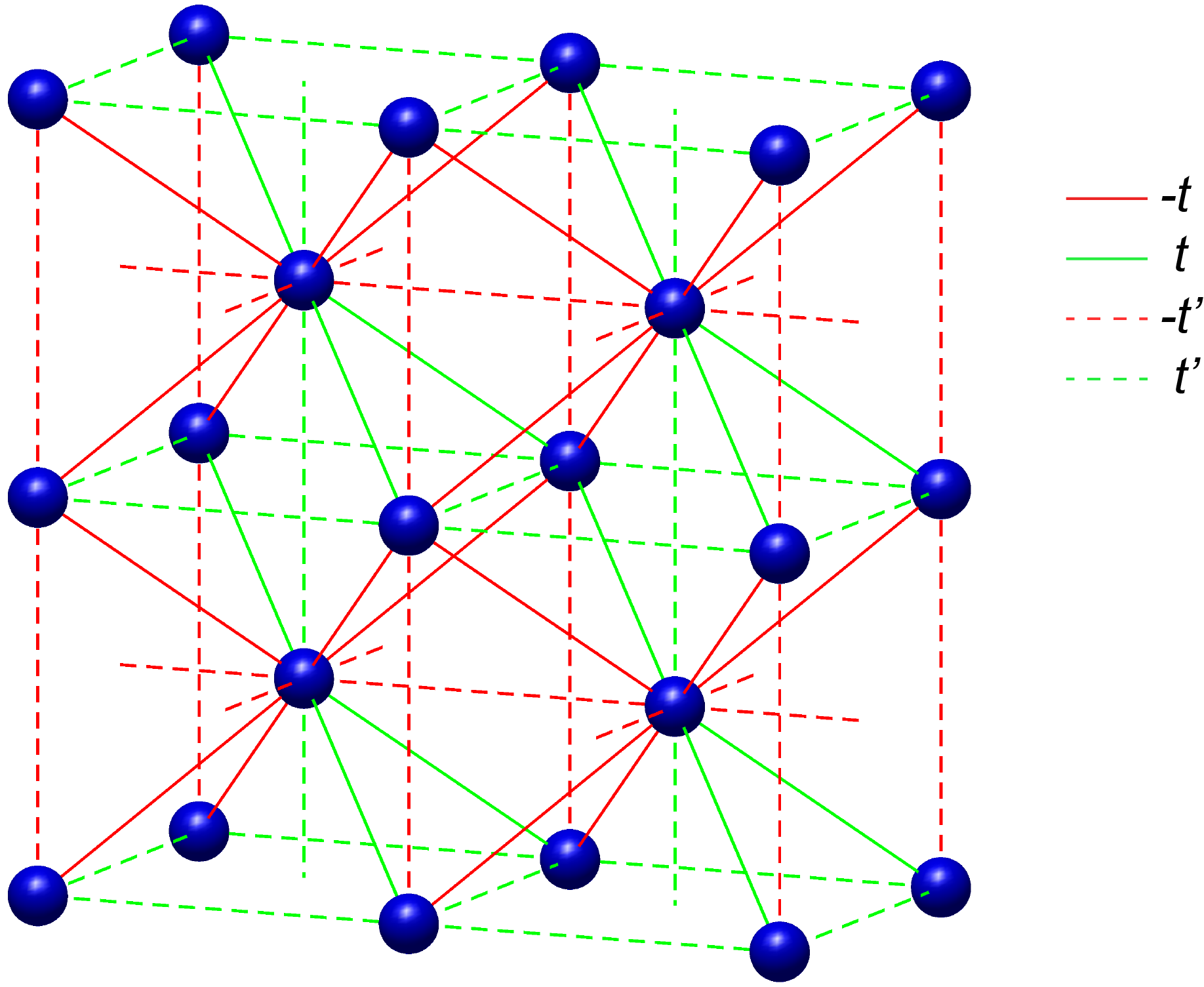}
       \caption{Crystal structure and hopping amplitudes in a Weyl semimetal obtained from a BCC alkali metal. The 
       nearest-neighbor hoppings, $t$ (solid lines), and next-nearest neighbor hoppings, $t'$ (dashed lines), entering the modified tight-binding model 
      of an alkali metal are displayed. For illustration purposes four unit cells of the lattice are shown.}
 \label{fig:fig0}
 \end{figure}

{\em Weyl semimetals from BCC metals.}
Our starting point is the band structure of simple alkali BCC metals. We approximate their lowest conduction band\cite{ashcroft}
by a simple tight-binding model\cite{harrison} containing a single-$s$ orbital per site, whose corresponding band dispersion reads:
$\epsilon({\bf k})=-8t \cos(k_x/2) \cos(k_y/2) \cos(k_z/2)-2t' (\cos(k_x) +\cos(k_y)+\cos(k_z)) $,
where the cubic lattice parameter is taken to unity, $a=1$.  $t, t'$ correspond to the nearest and next-nearest neighbor hopping amplitudes, 
respectively. 
Typically, we will have that $t'<t$ but we leave $t'/t$ as an adjustable parameter, which will be shown not 
to modify our main results qualitatively. At half-filling the Fermi energy is precisely at zero, $\epsilon_F=0$, since the model respects 
particle-hole symmetry. 

In order to transform a BCC metal into a Weyl semimetal we artificially modify the 
original hoppings of the tight-binding model which, as discussed below, could be achieved by applying external time and spatially-dependent 
electric fields. We find that the hopping pattern in the BCC lattice shown in Fig. \ref{fig:fig0} leads to a Weyl semimetal. The resulting modified Hamiltonian reads:
\begin{equation}
H({\bf k}) = f_x({\bf k}) \sigma_x+ f_y({\bf k}) \sigma_y +f_z({\bf k})\sigma_z,
\label{eq:ham}
\end{equation}
with: $f_x({\bf k})=-2 t[\cos({\bf k}\cdot {\bf a}_4)-\cos({\bf k}\cdot {\bf a}_1)+\cos({\bf k}\cdot {\bf a}_2) ]$, 
$f_y({\bf k})= 2 t \sin({\bf k}\cdot {\bf a}_3)$ and $f_z({\bf k})=-2 t' [\cos(k_x)+\cos(k_y)-\cos(k_z))$
with the relative positions between n.n. atoms given by: ${\bf a}_1={1 \over 2} (1,1,-1)$, ${\bf a}_2={1 \over 2} (-1,1,1)$,
${\bf a}_3={ 1 \over 2} (1,-1,1)$, and ${\bf a}_4={ 1 \over 2} (1,1,1)$. The Pauli matrices $\sigma_i$ describe the 
pseudospin associated with the two inequivalent sites arising from the modified hopping pattern described
by $t$ and $t'$.

\begin{figure}
   \centering
\includegraphics[width=4.25cm]{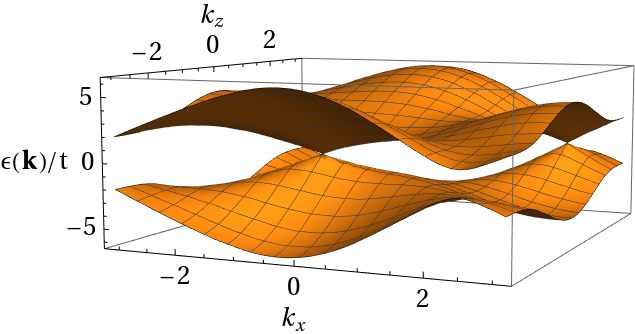}
\includegraphics[width=4.25cm]{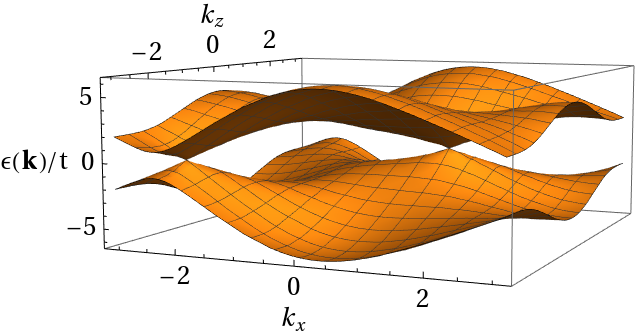}
\includegraphics[width=4.cm]{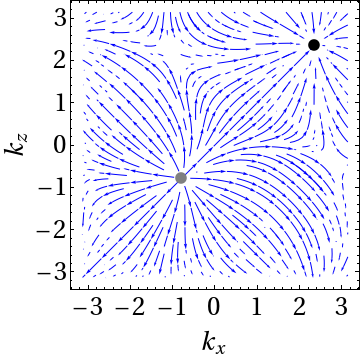}
\includegraphics[width=4.cm]{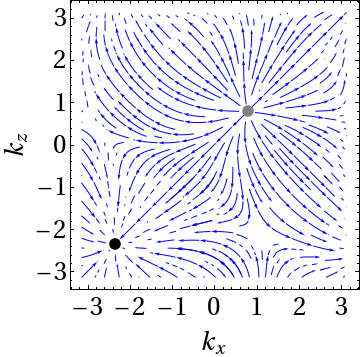}
       \caption{Weyl points and Berry curvatures in the Weyl semimetal generated from a BCC metal. Two Weyl points with opposite chiralities in the
       $k_x$-$k_z$ plane are shown in the top row for $k_y=-\pi/2$ (left) and $k_y=\pi/2$ (right). The corresponding projected Berry 
       curvatures are shown in the bottom row. Inward and outward flows of Berry curvature 
       indicate the presence of Weyl points with negative (black dot) and positive (grey dot) chiralities, respectively.}
 \label{fig:fig1}
 \end{figure}

The new hopping pattern transforms the original band of the BCC metal 
into two bands: $\epsilon^{\pm}({\bf k})=\pm \sqrt{ f_x({\bf k})^2 + f_y({\bf k})^2 + f_z({\bf k})^2} $, 
shown in Fig. \ref{fig:fig1}. We find that there are Weyl points, {\it i. e.}, ${\bf k}$-points in the three-dimensional momentum space  
at which two non-degenerate bands dispersing linearly around these points, touch at zero energy.  
Such accidental degeneracies are allowed by the fact that in three-dimensional solids there are three 
independent parameters: $k_x$, $k_y$, and $k_z$ that can be tuned to simultaneously satisfy the set of equations: 
$f_x({\bf k})=f_y({\bf k})=f_z({\bf k})=0$. This occurs at the Weyl points given in Table. \ref{table0}.

We have intentionally constructed a modified Hamiltonian (\ref{eq:ham}) for the BCC metal that preserves
 time-reversal symmetry (TRS) ($H({\bf k})=H^*(-{\bf k}))$, but breaks spatial inversion symmetry (IS) ($\sigma_x^{-1} H(-{\bf k})) \sigma_x \neq H({\bf k})$),
 in order to avoid the difficulties inherent to detecting Weyl fermions in materials with broken time-reversal symmetry. 
Heavy alkali metals such as Cs, which have significant spin-orbit coupling, are appropriate candidates for 
generating a BCC Weyl semimetal since non-degenerate bands  (except 
at time-reversal inversion symmetry momenta) are expected once inversion symmetry is broken.

{\em Topological properties: Weyl nodes, chiralities and Berry curvature. }
The generated Weyl semimetal has characteristic topological properties
associated with Weyl nodes. The Nielsen-Nimoniya theorem \cite{nielsen1983} imposes that Weyl points in a lattice 
should always occur in pairs of opposite charges. In a TRS system, the chiralities,  $\chi= |\nu_{\alpha\beta}| $, where $\nu_{\alpha\beta}={\partial f_\beta \over \partial k_\alpha}$,  ($\alpha,\beta=x,y,z$), 
of the Weyl points at ${\bf k}_W$ and $-{\bf k}_W$ are equal. This is because TRS implies: $f_x(-{\bf k})=f_x({\bf k}), f_y(-{\bf k})=-f_y({\bf k}), f_z(-{\bf k})=f_z({\bf k})$, leaving the 
products $\nu_{xi} \nu_{yj} \nu_{zk} $ entering the chirality invariant under the interchange ${\bf k}_W \rightarrow - {\bf k}_W$. 
Hence, the minimum number of Weyl points 
in a TRS system is four. In our calculations we indeed find four Weyl points (see Table \ref{table0}) whose location is independent of $t'/t$.

 \begin{table}[b]
\caption{Weyl points found in BCC metals described with the modified tight-binding hoppings, $t$ and $t'$, 
of Fig. \ref{fig:fig0}. The position of the Weyl points in {\bf k}-space (${\bf k}_W$) and their chirality ($\chi$) is independent of $t'/t$. }
\label{table0}
\begin{tabular}{ll}
 ${\bf k}_W$ & $\chi$   \\
\hline
$(\pi/4,\pi/2,\pi/4)$   & +1    \\
$(-\pi/4, -\pi/2, -\pi/4)$   &  +1    \\
$(3\pi/4,-\pi/2,3 \pi/4) $  &  -1   \\
$(- 3 \pi/4,\pi/2, -3 \pi/4)$   &  -1  \\
\hline
\end{tabular}
\end{table}

\begin{figure*}
   \centering    
\includegraphics[width=5cm]{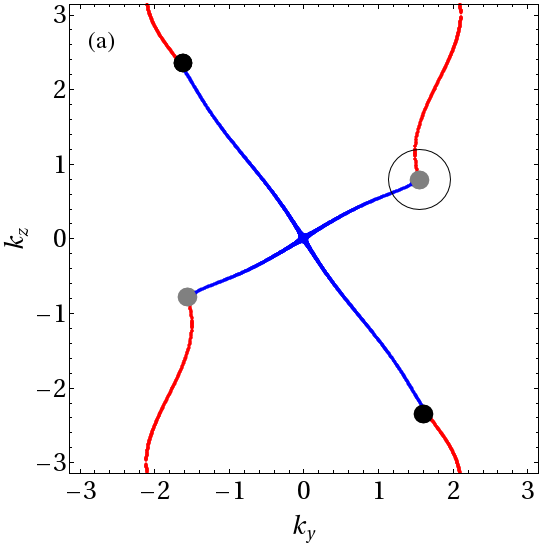}
\includegraphics[width=5cm]{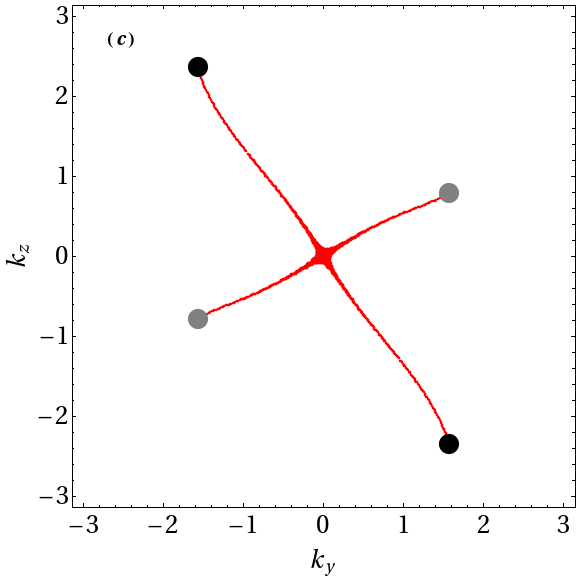}
\includegraphics[width=5cm]{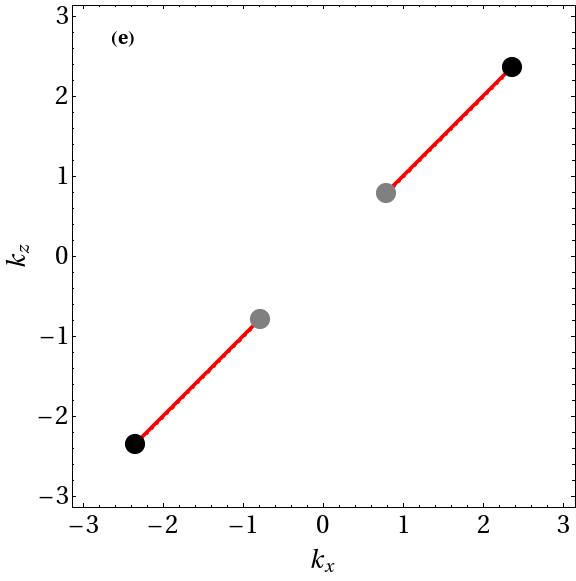}
\includegraphics[width=5.cm]{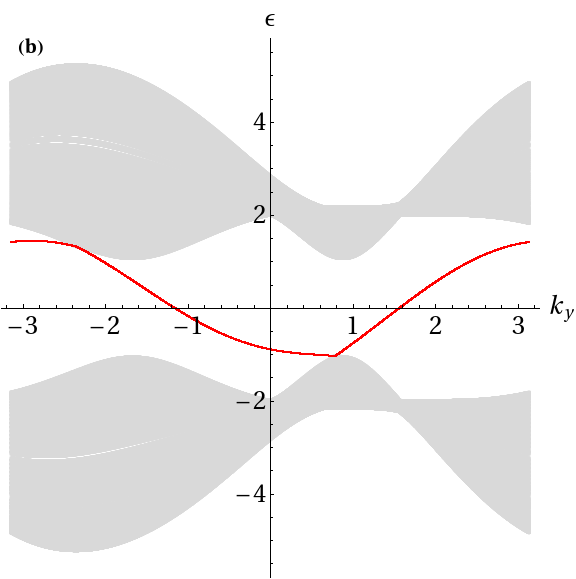}
\includegraphics[width=5.cm]{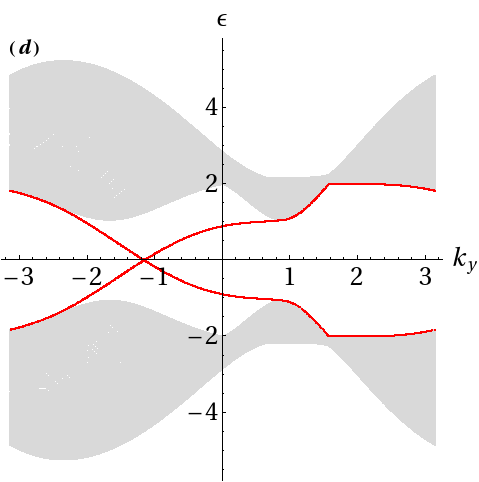}
\includegraphics[width=5.cm]{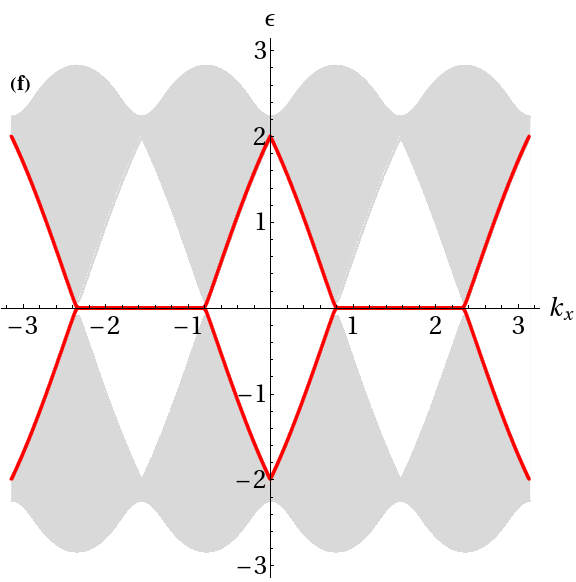}
       \caption{Topological surface bands and Fermi arcs in BCC Weyl semimetals. (a) Calculated Fermi arcs on the front (blue line) and back (red line) (100) surfaces 
       of a slab with an odd number of layers. (b) Bulk (gray continuum) and surface band dispersions (red line) projected along the $k_y$ direction at $k_z=\pi/2$ for the (100) slab of (a). 
       (c), (d) same as (a), (b) respectively, but for a slab with an even number of layers. (e) Calculated Fermi arcs connecting Weyl points 
       projected on the (010) surface of a slab with an even number of layers. (f) Bulk and surface band dispersions along the $k_y=k_x$ 
       direction for the (010) slab of (e). Note that the Fermi arcs on opposite sides of the slab coincide in the cases (c) and (e) for an even number of layers 
       but not in (a) with an odd number of layers. The black (gray) dots represent the projection of the Weyl points on the different surface orientations 
       with negative (positive) chiralities. The gray circle describes the edge of a cylinder along the $x$-direction cutting the (100) surfaces
       of the Brillouin zone which encloses a single Weyl point with positive chirality. Each edge of the cylinder sustains a topologically protected 
       one-dimensional edge state contributing a state to the Fermi arc. The surface bands connecting the valence and conduction bulk bands lead 
       to gapless metallic surfaces corroborating their topological nature.}
 \label{fig:fig2}
 \end{figure*}

Weyl points can be considered hedgehogs of Berry curvature in momentum space\cite{volovik2013}, behaving analogously to magnetic monopoles in real space.
This is evident from calculations of the Berry curvature\cite{berry1984}: ${\bf \Omega}({\bf k})={\bf \nabla}_{\bf k} \times \langle u({\bf k}) |i {\bf \nabla}_{\bf k} | u({\bf k}) \rangle $, 
where $|u({\bf k}) \rangle $ is the eigenfunction of the occupied band. Figure \ref{fig:fig1} shows how Weyl points act as sources and drains of Berry curvature 
projected on the $k_x$-$k_z$ plane in analogy with the magnetic field around magnetic monopoles in real space. Weyl points have a quantized Berry flux\cite{xiao2010} given by:
${1 \over 2 \pi } \iint_{S^2} {\bf \Omega} ({\bf k}) d {\bf k}=\chi$, so the chirality can be interpreted as a topological Chern number of the Weyl node. We only find two possible chiralities of the Weyl points: $\chi=\pm 1 $, in the band structure considered here. Due to TRS the Berry curvature satisfies\cite{haldane2004,xiao2010}: ${\bf \Omega}(-{\bf k})=-{\bf \Omega}({\bf k})$   
implying that there is no anomalous quantum Hall contribution to the Hall conductivity.
However, an anomalous contribution to the electron velocity, not expected in conventional alkali metals, would arise under a weak uniform electric field\cite{xiao2010}:
 ${\bf v}_n({\bf k})= {\partial \epsilon_n({\bf k}) \over \hbar \partial {\bf k} } - {e \over \hbar} {\bf E} \times {\bf \Omega}({\bf k})$. 

To further investigate the topological properties of the analyzed structure, we turn now to study the effect of adding an external staggered mass term, $m\sigma_z$, to our Hamiltonian (\ref{eq:ham}). Our calculations show that the Weyl nodes move around in ${\bf k}$-space
until pairs of opposite chiralities meet at certain momenta (see details in the Suppl. Information, \cite{suppl}). The observed 
robustness of the Weyl points to the staggered mass term is a consequence of the breaking of spatial inversion symmetry
by our Hamiltonian. In particular, a Weyl semimetal with four Weyl nodes is found for staggered masses in the range: $-4 (t'/ t) < m \lesssim 2.83 (t'/ t) $. We find that a  trivial band insulator is realized when $m < -4(t'/t)$, whereas a three-dimensional (3D) topological insulator with topological surface states arises
for $m>2.83 (t'/t)$ as we discuss below. Hence, the Weyl semimetal generated here by breaking inversion symmetry is an intermediate metallic state between a 
band insulator and a three-dimensional topological insulator \cite{murakami2017}.   


{\em Topological surface bands and Fermi arcs.}
In gapped topological matter such as topological insulators\cite{kane2010} the bulk-boundary correspondence gives rise to gapless fermions
on the surface of the system\cite{volovik2013}. Similarly, in topological gapless materials such as Weyl semimetals, a crucial signature of Weyl fermions 
is the presence of gapless topological surface states. These surface states lead to exotic unclosed Fermi arcs\cite{hasan2017} in the surface Brillouin zone.
We have computed the surface states on slabs which are cut along different crystal directions of the BCC Weyl semimetal. 
In Fig. \ref{fig:fig2} we show the Fermi arcs and surface bands on the (100) and (010) surfaces. We predict Fermi arcs
connecting the Weyl points with opposite chiralities at the surfaces of both sides of the slab. The Fermi arc shapes change
depending on the termination of the surface {\it i.e.} whether the number of layers in the slab direction is even or odd. 
Fermi arcs can be detected not only by ARPES experiments but also through standard  quantum oscillation
experiments in spite of the arcs being open. This is because under a magnetic field, electrons circulating along a Fermi arc on one side of the slab 
can traverse the bulk via the Weyl points to the Fermi arc on the opposite side of the slab leading to closed magnetic orbits.\cite{kimchi2014} 
These unusual closed orbits can be detected in quantum oscillatory phenomena and should display topological signatures of the Weyl semimetal.\cite{moll2016}  

We focus first on the results for the (100) surfaces shown in Fig. \ref{fig:fig2} (a)-(d). The structure of the BCC crystal 
shown in Fig. \ref{fig:fig0} contains two types of atoms A and B. For a cut of the crystal
with an even number of layers both surfaces contain atoms of the same type. On the other hand, for a cut involving 
an even number of layers in the $x$-direction we have that 
one of the surfaces is formed by A(B)-sublattice atoms whereas the opposite side is formed
by atoms of the B(A)-sublattice. In Fig. \ref{fig:fig2} (a)-(b) we show the Fermi arcs and surface bands of the (100) surfaces
of a slab with an odd number of layers along the $x$-direction. The Fermi arcs at the front and at the back surfaces are found to 
have very different shapes as shown in Fig. \ref{fig:fig2}(a). This is in contrast to even number terminations in which the 
Fermi arcs on opposite surfaces coincide.  The surface bands connect the
valence and conduction bands closing the bulk gap leading to metallic surfaces. 

On the other hand, our calculations for the (010) surface of Fig. \ref{fig:fig2}(e) show how two straight lines connect the two 
pairs of Weyl points with opposite chiralities along the diagonal ($k_z=k_x$ direction) of the $k_x$-$k_z$ surface. 
These open and disjoint Fermi lines result from the flat surface band located right at the Fermi energy shown in
Fig. \ref{fig:fig2}(f). Note that this flat band is similar to the one-dimensional surface band connecting two Dirac points
in "spinless" graphene nanoribbons\cite{schnyder2016}.

The topological protection of the surface states is guaranteed by the bulk-boundary correspondence. 
This can be explicitly illustrated by considering, for example, the (100) face by taking a cylinder  
which extends along the $x$-direction of the whole Brillouin zone and encloses a single Weyl point\cite{hasan2017}. For example, 
the cylinder taken in Fig. \ref{fig:fig2} (a) has a Chern number of +1 (since it contains a Weyl point with $\chi=+1$) 
giving rise to topologically protected edge states at the one-dimensional edges of the tube. Hence, this two-dimensional 
slice of the Brillouin zone behaves as a quantum Hall insulator with chiral states protected by a non-zero Chern number. 
The two edge states of the cylinder make slices of the Fermi arcs on the two (100) surfaces of the slab. 
By varying the radius of the cylinder the full Fermi arcs can be reconstructed.   

The character of the different insulating phases arising under the staggered mass term can be obtained by calculating surface states. 
We find that for a positive staggered mass, $m>2.83 (t'/t)$, the Fermi arcs evolve into a closed Fermi surface at the Brillouin zone edge 
consistent with a 3D topological insulator whereas for negative staggered masses, $m<-4 (t'/t)$, the surface
is gapped signalling a conventional band insulator (see Suppl. information \cite{suppl}). Hence, we conclude that the Weyl semimetal
is an intermediate semimetallic state arising in a transition between a band insulator and a 3D topological insulator.

{\em Modification of the hopping amplitudes in a tight-binding model for BCC alkali metals.}
The actual implementation of the proposed hopping pattern can benefit from previous works which aim at engineering
of couplings using external applied fields. Linearly and circularly polarized light irradiation has been proposed for engineering spin exchange couplings in Mott insulators \cite{balents2018},
generate topological properties in graphene\cite{kitagawa2011,grushin2014}, create Floquet topological insulators\cite{kitagawa2011} and/or induce 3D topological 
bands in trivial band insulators.\cite{galitski2011,galitski2013} On the other hand, photo-assisted tunneling 
provides a framework for designing optical lattices\cite{buljan2015,grushin2018,eckardt2017,porras2015,bloch2013,ketterle2013} with the
desired properties. In this context, it is important to realize that the hopping pattern of Fig. \ref{fig:fig0} involves changing {\it only} the signs of 
the {\it real} hopping amplitudes entering the tight-binding model of alkali metals. Obtaining this type of hopping patterns 
has been shown to be possible by applying a uniform constant field combined with a uniform
on-site AC potential to {\it non-frustrated} simple cubic lattices. However, a succesful application of this approach to the BCC lattice is 
immediately precluded by the large connectivity of the lattice. In essence, this large connectivity leads to a large set of nonlinear
equations that cannot be solved simultaneously due to the small number of available parameters describing the external fields. 
In addition, we have numerically checked that even including arbitrary site-dependent phases does not provide enough degrees of freedom
to obtain the hoppings of Fig. \ref{fig:fig0}. Therefore, we conclude that a higher degree of flexibility in the applied external fields
(such as using combinations of site-dependent AC potentials) would be needed in order to generate the desired hopping pattern.

{\em Conclusions.} 
In summary, we have discovered a Weyl semimetal with an underlying BCC crystal structure 
obtained by a simple modification of the hopping pattern describing conventional alkali metals. The band 
structure generated in this way contains the minimum four Weyl 
nodes required by time-reversal symmetry. 
Fermi arcs connecting Weyl points projected on different surfaces arise 
from surface states topologically protected by non-zero Chern 
numbers of $\pm 1$.  
Our work can contribute to the further understanding of Weyl phenomena by providing 
simple material platforms hosting energy-isolated Weyl nodes. 


\section*{Acknowledgements.}
We acknowledge financial support from (MAT2015-66128-R) MINECO/FEDER, Uni\'on Europea.

\end{document}